\documentclass[prl,twocolumn,showpacs,floatfix]{revtex4}
\usepackage{graphicx,amsmath,amssymb,mathptm}
\begin{document}
\title{Complete Condensation in a Zero Range Process on Scale-Free Networks}
\author{Jae Dong Noh, G. M. Shim, and Hoyun Lee}
\affiliation{Department of Physics, Chungnam National University, Daejeon
305-764, Korea}
\date{\today}

\begin{abstract}
We study a zero range process on scale-free networks in order to
investigate how network structure influences particle dynamics.
The zero range process is defined with the rate $p(n) = n^\delta$ 
at which particles hop out of nodes with $n$ particles.
We show analytically that a complete condensation occurs 
when $\delta \leq \delta_c \equiv 1/(\gamma-1)$ where $\gamma$ is the degree
distribution exponent of the underlying networks. In the
complete condensation, those nodes whose degree is higher than a threshold 
are occupied by macroscopic numbers of particles, while the
other nodes are occupied by negligible numbers of particles. We also show
numerically that the relaxation time follows a 
power-law scaling $\tau \sim L^z$ with the network size $L$ and 
a dynamic exponent $z$ in the condensed phase.
\end{abstract}

\pacs{89.75.Hc, 05.20.-y, 02.50.Ey}
\maketitle

Network systems have been attracting much interest in recent
years~\cite{Watts98,Albert02,Dorogovtsev02,Newman03}. 
A network is a simplified description for complex systems, such as the
Internet, with a set of nodes interconnected with links.
Various networks have been introduced and studied in wide areas of 
science, many of which share the common 
feature of the power-law degree distribution $P_{\text{deg.}}(k) \sim 
k^{-\gamma}$~\cite{Albert02}. A degree of a node denotes the number
of links attached to it. Such a network is called a scale-free~(SF) 
network.

The SF network has the broad degree distribution, which leads to many 
intriguing properties of dynamical systems defined on it.
One of those is a centralization. For instance,
a random walker on a network tends to visit more frequently 
those nodes with higher degree~\cite{Noh04}. 
Higher degree nodes usually have large values of the load or betweenness
centrality, which is a measure of a traffic along shortest pathways~\cite{Newman01,Goh02}. 
Whether under stochastic or deterministic dynamics, 
particles tend to concentrate on high degree nodes.
Consequently, one expects that a particle interaction would play an
important role in dynamic processes.
However, only a little attention has been paid
to interacting dynamical systems on SF networks~\cite{Holme}.

In this work we study a zero range process~(ZRP)~\cite{Evans00,Grosskinsky03}
on SF networks. The ZRP is a model for condensation in interacting stochastic 
particle systems. It has been widely studied since it has a lot of 
applications such as traffic flows and its stationary state properties are
solvable. In this study, we investigate the interplay between particle 
interaction and underlying network structure.

Consider a ZRP for $N$ particles on a network of $L$ nodes. 
We mainly focus on the limiting case where $N,L\rightarrow\infty$  
with fixed density $\rho=N/L$.
Each node $i=1,\ldots,L$ may be occupied by any number of particles. 
We denote the occupation number at each node $i$ with $n_i$ and 
a microscopic state with ${\mathbf{n}} = (n_1,n_2,\ldots,n_L)$.
In dynamics, only a single particle is allowed to jump out of
each node during a unit time interval~\cite{comment1,Evans04}, 
and the jumping rate $p_i$ out of a node $i$ depends only on the 
occupation number at the source node. For simplicity, we assume that 
the jumping rate is given by the same function of the occupation number 
for all nodes, that is, $p_i = p(n)$. 
A jumping particle out of $i$ then hops to a target node $j$ 
according to a prescribed transition probability denoted by $T_{j\leftarrow i}$.

The model incorporates particle interactions with the hopping 
rate $p(n)$. When $p(n)\propto n$, particles are noninteracting and move 
independently. 
A repulsive interaction is imposed when $p(n)$ increases faster than $n$.
An attractive interaction is imposed when $p(n)$ 
increases sublinearly~[weak attraction] or deceases~[strong
attraction] with $n$.
The ZRP may also model a transport process, where
e.g., data packets are delivered from one node to another. 
In that context, $p(n)$ corresponds to the amount of packets
that can be handled per unit time by each node with $n$ packets,
that is, the jumping rate function may represent a transport capacity 
of each node.

The ZRP is solvable as far as stationary state properties are concerned.
We give a brief review in the following.
For detailed review, we refer readers to
Refs.~\cite{Evans00,Grosskinsky03}. 
Let $P_{\text{stat.}}(\mathbf{n})$ be the stationary state probability 
distribution. Interestingly, it has the product form~\cite{Evans00}:
$$
P_{\text{stat.}}(\mathbf{n}) = \frac{1}{Z(L,N)} \prod_{i=1}^L f_i (n_i) \ ,
$$
where $f_i(n) = \prod_{l=1}^n \left[{\omega_i}/{p(l)}\right]$
for $n>0$ and $f_i(0) = 1$ and $Z(L,N)$ is a normalization constant.
Here, $\{\omega_i\}$ is the stationary state probability
distribution of a single particle or a random walker moving on the 
network with the transition probability $\{T_{i\leftarrow j}\}$.
The mean occupation number $m_i = \langle n_i \rangle = 
\sum_{\mathbf{n}}n_i P_{\text{stat.}}(\mathbf{n})$ is given by
$ m_i  = z ({\partial}/{\partial z}) \ln F_i (z)$,
where
\begin{equation}\label{eq:F}
F_i(z) \equiv \sum_{n=0}^\infty \ z^n f_i(n) \ .
\end{equation}
The series is defined within the interval $|z|<z_c$, where $z_c$ is the radius
of convergence.
The fugacity $z$ should be determined self-consistently from
$\rho =  (\sum_{i=1}^L m_i)/L$.

In the periodic lattices with
$\omega_i=\mbox{const.}$, the {\em condensation} occurs 
when $p(n)$ decays faster than $p_c(n)\equiv 1+2/n$ and $\rho>\rho_c$ with
a critical density $\rho_c$~\cite{Evans00}. 
In the condensed state a node is occupied by macroscopic 
numbers of particles.
It will turn out that the condensation occurs in a different way in SF
networks. 

Now we take a SF network of $L$ nodes with 
the power-law degree distribution $P_{\text{deg.}}(k_0 \leq k\leq k_{max}) 
\sim k^{-\gamma}$. The degree distribution exponent is $\gamma$ and 
$k_0$ is a constant. We only consider the network with $\gamma>2$ so that 
the mean degree $\overline{k}=\int dk k P_{\text{deg.}}(k)$ is finite.
The node with the maximum degree $k_{max}$ is called the {\em hub}.
For a  given $P_{\text{deg.}}(k)$,
the maximum degree scales as $k_{max} \sim L^{1/(\gamma-1)}$, which
results from $\int_{k_{max}}^\infty P_{\text{deg.}}(k)dk = 
\mathcal{O}(1/L)$.

We adopt the following jumping rate function
\begin{equation}
p(n) = n^\delta \ ,
\end{equation}
where the parameter $\delta$ controls the interaction strength or
the transport capacity.
A particle jumping out of a node $i$ is allowed to hop 
to one of the neighboring nodes of $i$ selected randomly.
That is, the transition probability
is given by $T_{j\leftarrow i}= 1/k_i$ if $i$ and
$j$ are connected and $T_{j\leftarrow i}=0$ otherwise, where $k_i$ denotes 
the degree of $i$. 

When there is only a single particle, the ZRP reduces to random walks
with the transition probability $\{T_{i\leftarrow j}\}$.
The random walk problem is studied in Ref.~\cite{Noh04}, where it is 
shown that the stationary state distribution is given by $\omega_i = k_i / E $ 
with $E=\sum_j k_j$.
One then obtains 
$f_i(n) = \prod_{l=1}^n [ \omega_i / p(l)] = (k_i/E)^n / (n!)^\delta$.
Hence the function $F_i(z)$ in Eq.~(\ref{eq:F}) takes the form
$F_i(z) = \mathcal{F}_\delta(x=zk_i/E)$ with
\begin{equation}\label{eq:Fx}
\mathcal{F}_\delta(x) = \sum_{n=0}^\infty \frac{x^n}{(n!)^\delta} \ ,
\end{equation}
and the mean occupation number is given by
\begin{equation}\label{eq:nx}
m_i  =  \left. x \frac{\partial}{\partial x}
                      \ln \mathcal{F}_\delta(x) \right|_{x= x_i = zk_i/E} \ .
\end{equation}
The parameter $x_i\equiv zk_i/E$ will be called the effective fugacity.
The higher the degree is, the larger the effective fugacity is.
Note that the particle distribution is determined solely by 
the degree distribution.
The mean occupation number is parameterized
with the fugacity $z$, whose value has to be fixed from 
$\rho = (\sum_{i} m_i)/L$. 

First of all, we derive the mean occupation number distribution in the
simplest case with $\delta=0$, where $\mathcal{F}_0(x) = 1/(1-x)$
for $x<1$ and $ m_i  = x_i / (1-x_i)$. 
The actual occupation number is determined after solving the relation 
$\rho = L^{-1}\sum_{i=1}^L {x_i}/{(1-x_i)}$.
In the $L\rightarrow\infty$ limit, one might replace the summation with the
integral. However, one has to be cautious since 
$m$ as a function of $x$ is singular at $x=1$. The hub has the largest
effective fugacity, hence one may safely decompose the
particle density into two terms as $\rho =  \rho_s + \rho_n$ for 
the hub and the other nodes, respectively. They are given by
\begin{eqnarray*}
\rho_s &=& \frac{m_{hub}}{L} = 
\frac{1}{L}\frac{zk_{max}/E}{1-zk_{max}/E} \\
\rho_n &=& \int_{k_0}^{k_{max}} dk P_{\text{deg.}}(k) \frac{zk /
E}{1-zk /E} = \sum_{l=1}^\infty \overline{ k^l}
\left(\frac{z}{E}\right)^l 
\end{eqnarray*}
with $\overline{k^l}= \int_{k_0}^{k_{max}} dk k^l P_{\text{deg.}}(k)$.
Using the power-law degree distribution, one obtains that
$\overline{k^l} \sim k_{max}^{l+1-\gamma}/l$ for large $l$. Hence the
total density at all nodes but the hub is given by
$\rho_n \sim  k_{max}^{-(\gamma-1)} | \ln (1-zk_{max}/E )|$.
One can eliminate the fugacity $z$ from the relation for $\rho_s$, which
yields that $\rho_n \sim k_{max}^{-(\gamma-1)} \ln (1+m_{hub})$. 
Note that $k_{max} \sim L^{1/(\gamma-1)}$ and that
$m_{hub}\leq \rho L$. It implies that $\rho_n$ vanishes as 
$\rho_n\sim L^{-1}\ln L$ in the $L\rightarrow\infty$ limit.
Consequently, the whole fraction of particles is condensed into the hub 
with $m_{hub}/L = \rho$.

The analysis shows that a {\em complete condensation} occurs at $\delta=0$, 
where the whole fraction of particles is concentrated at the hub.
Moreover, the condensation takes place at any
finite value of $\rho$. With these two features the complete condensation 
contrasts well with the usual condensation occurring in homogeneous networks.

In the non-interacting case~($\delta=1$), it is trivial to show that $m_i
= \omega_iN = \rho k_i / \overline{k}$ with the mean degree $\overline{k}$. 
The hub also has the largest occupancy,
but it scales sublinearly with $L$ as $m_{hub} \sim k_{max} \sim 
L^{1/(\gamma-1)}$.
So the condensation does not occur at $\delta=1$. 
It will be shown that there exists a critical value of $\delta$ 
for the condensation.

For the general cases with $\delta>0$, we need to figure out the
property of $\mathcal{F}_\delta(x)$ in Eq.~(\ref{eq:Fx}).
The radius of convergence is infinite and it is analytic for all values 
of $x$.
Since a closed form expression is not available, we try to find an
approximation.
For large $x$, the series is dominated by the terms near $n=n_0$ at which 
the summand $x^n / (n!)^\delta \equiv e^{g(n)}$ is maximum. 
Using the Stirling formula and
treating $n$ as a continuous variable, we obtain $n_0\simeq
x^{1/\delta}$ from $g'(n_0)=0$.
Then, the series is approximated as
$\mathcal{F}_\delta(x) \simeq \sum_{n=0}^\infty \exp[ g(n_0) + 
g''(n_0) (n-n_0)^2/2 + \cdots]$.
Replacing the summation with the integral, we obtain that 
\begin{equation}\label{eq:Fap}
\mathcal{F}_\delta(x) \simeq \frac{1}{\sqrt{\delta}}  
\left( {2\pi x^{1/\delta}} \right)^{(1-\delta)/2} e^{ \delta
x^{1/\delta}} \ .
\end{equation}
Note that the approximation coincides with the exact form
$\mathcal{F}_1(x) = e^x$ at $\delta=1$.
We have checked numerically that the approximation is also good for 
$\delta\neq 1$ unless $x$ is too close to zero. 
For small $x$, we can approximate the series with a few lowest order 
terms.

Within the approximation, the mean occupation number 
is given by
\begin{equation}\label{eq:m_x}
m_i \simeq \left\{
 \begin{array}{cl}
   x_i & \mbox{for $x_i\ll 1$}, \\
   x_i^{1/\delta} & \mbox{for $x_i\gtrsim 1$} \ .
 \end{array}
 \right.
\end{equation}
The self-consistency relation $\rho=\sum_i m_i / L$ requires that 
$x_{hub}=zk_{max}/E$ should be 
larger than 1.
So we get $m_{hub} = x_{hub}^{1/\delta}$ or 
$z = E m_{hub}^\delta/ k_{max}$. Then, the effective
fugacity for other nodes is written as $x_i = k_i / k_c$ where $k_c$ is 
the crossover degree defined as
\begin{equation}\label{eq:kc}
k_c = \frac{k_{max}}{m_{hub}^\delta}\ .
\end{equation}
The effective fugacity ranges from $x_{min} =  (k_0/k_c)$.

The self-consistency relation $\rho=\sum_i m_i/L$ is used to find the value
of $k_c$.
It takes a different form
depending on whether $x_{min}$ is larger than $1$ or not.
First, let us assume that $x_{min}\gtrsim 1$, or equivalently, 
$k_0 \gtrsim k_c$.
Then, $m_i\simeq x_i^{1/\delta}$  for all nodes,
and the self-consistency relation becomes as 
$\rho = \overline{k^{1/\delta}} / k_c^{1/\delta}$.
It is compatible with the assumption $k_0 \gtrsim k_c$ 
only when $\overline{k^{1/\delta}}$ is non-divergent. 
Note that $\overline{k^{1/\delta}} = \int dk k^{1/\delta} P_{\text{deg.}}(k) 
\sim \int_{k_0}^{k_{max}} dk k^{1/\delta - \gamma}$.
It is non-divergent when $\delta > \delta_c$ where
\begin{equation}
\delta_c = \frac{1}{\gamma-1}\ .
\end{equation}
In that case, the mean occupation number is given by
$m_{i} = c k_{i}^{1/\delta}$
with a constant $c={\rho}/{\overline{k^{1/\delta}}}$.
The hub has the largest mean occupation number, 
but it scales sublinearly with $L$
as $m_{hub} \sim k_{max}^{1/\delta} \sim L^{\delta_c /\delta}$ with
$\delta_c/\delta<1$. Therefore we conclude that the system does not
show the condensation for $\delta > \delta_c$.

Let us assume the opposite situation where $x_{min}\ll 1$ or $k_0 \ll k_c$.
Then, the self-consistency relation is decomposed into two parts 
as $\rho = \rho_n + \rho_s$ for nodes with $k_i < k_c $ and $k_i > k_c$,
respectively. According to Eq.~(\ref{eq:m_x}), they are given by
$\rho_n = \sum_{k_i < k_c} x_i / L$
and $\rho_s = \sum_{k_i>k_c} x_i^{1/\delta} / L$.
The summands have no singularity, and one can write in the
$L\rightarrow\infty$ limit as
\begin{eqnarray*}
\rho_n &=& k_c^{-1} \int_{k_0}^{k_c} dk k P_{\text{deg.}}(k) \\
\rho_s &=& k_c^{-1/\delta} \int_{k_c}^{k_{max}} dk k^{1/\delta} 
P_{\text{deg.}}(k) \ .
\end{eqnarray*}
Note that $\int_{k_0}^{k_c} dk k P_{\text{deg.}}(k)$ is smaller than 
the mean degree $\overline{k}$, hence is finite.
So $\rho_n$ vanishes as $\sim k_c^{-1}$, which yields that $\rho_s = \rho$ and
$k_c^{1/\delta} = \int_{k_c}^{k_{max}} dk k^{1/\delta}P_{\text{deg.}}(k)
/\rho \sim \int_{k_c}^{k_{max}} dk k^{1/\delta - \gamma}$.
It is compatible with the assumption that $k_c\gg k_0$ only when
the integral is divergent, which requires that
$\delta = \delta_c$ or $\delta < \delta_c $.
In either case, we obtain that $k_c \sim [\ln k_{max}]^{\delta_c}$ or
$k_c \sim k_{max}^{1-\delta/\delta_c}$, respectively.
The occupation number distribution is then easily found using
Eqs.~(\ref{eq:m_x}) and (\ref{eq:kc}), 
which is summarized in Table~\ref{table1}.
For $\delta\leq \delta_c$, the mean occupation number, 
as a function of the degree, satisfies the scaling form 
\begin{equation}\label{m_scale}
m_k = G_\delta(k/k_c) \ ,
\end{equation}
where the scaling function behaves as $G_\delta(y\ll 1) \sim y$ and 
$G_\delta(y \gtrsim 1)\sim y^{1/\delta}$.

\begin{table}[t]
\caption{Scaling of the crossover degree scale $k_c$, 
the mean occupation number, and $m_{hub}$.}
\label{table1}
\begin{ruledtabular}
\begin{tabular}{ccccc}
$\delta$  &   $k_c$  &  $m_{k<k_c}$ & $m_{k>k_c}$ & $m_{hub}$ \\ \hline
$\delta > \delta_c$  &  ---   &   --- & $ k^{1/\delta}$ &
                        $\mathcal{O}(L^{\delta_c/\delta})$ \\
$\delta = \delta_c$  &  $[\ln k_{max}]^{\delta_c}$ & 
                        $k / [\ln k_{max}]^{\delta_c}$  & 
                        $k^{1/\delta_c} / \ln k_{max}$  &
                        $\mathcal{O}(L/\ln L)$\\
$\delta < \delta_c$  &  $k_{max}^{1-\delta/\delta_c}$ & 
                        $k / k_{max}^{1-\delta/\delta_c} $  &
                        $k^{1/\delta} /  k_{max}^{1/\delta-1/\delta_c}$ &
                        $\mathcal{O}(L)$ \\
$\delta = 0$         &  $k_{max}$ & $k/(k_{max}-k)$ & --- & $\rho L$
\end{tabular}
\end{ruledtabular}
\end{table}

The maximum degree scales as $k_{max} \sim L^{\delta_c}$, which gives 
$k_c \sim L^{\delta_c-\delta}$ for $\delta<\delta_c$.
Using this in Table~\ref{table1} for $\delta<\delta_c$, 
we find that the nodes with $k >k_c$ 
are occupied by macroscopic numbers of particles as
$m_{k>k_c} \sim L (k/k_{max})^{1/\delta} = \mathcal{O}(L)$. 
Furthermore, the whole fraction of particles is distributed on those
nodes, which is clear from $\rho_n \sim k_c^{-1} \sim L^{-(\delta_c-\delta)}$.
Therefore, we conclude that the system displays the complete condensation for
$\delta < \delta_c$.
At the marginal case $\delta=\delta_c$, the 
complete condensation also occurs with a logarithmic correction~(see
Table~\ref{table1}).
Note that the case with $\delta=0$ corresponds to that with $k_c = k_{max}$
so that whole fraction of particles are condensed on to the hub. The results
at $\delta=0$ are also summarized in Table~\ref{table1}.

\begin{figure}[t]
\includegraphics*[width=\columnwidth]{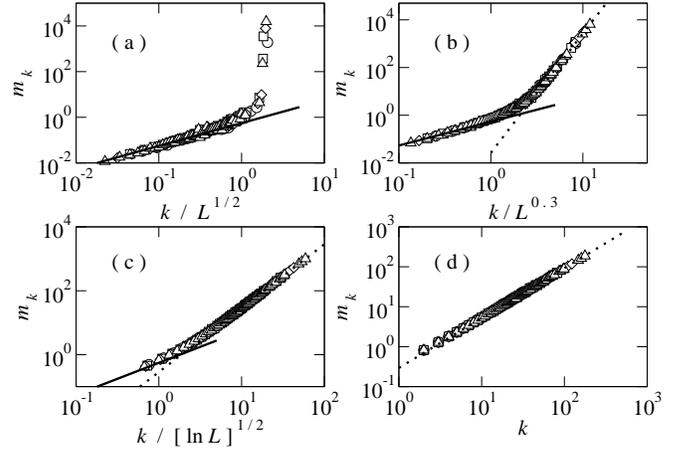}
\caption{Scaling of the mean occupation number at $\delta=0.0$ in (a), 
$0.2$ in (b), $0.5$ in (c), and $0.8$ in (d). Data from different network
sizes are distinguished with different symbols~($\circ$ for $L=1000$, 
$\Box$ for $L=2000$, $\diamond$ for $L=4000$, and $\triangle$ for
$L=8000$). The slope of  solid lines is 1 and that of dotted lines
is $1/\delta$.}
\label{fig1}
\end{figure}

The analytic results are confirmed by Monte Carlo
simulations on the archetypical  
Barab\'asi-Albert~(BA) network~\cite{BAnet} with $\gamma=3$ and 
$\langle k\rangle=4$. It has $\delta_c = 1/2$. 
The network sizes are $L=1000, 2000, 4000, 8000$ and the total density
of particles is fixed to $\rho=2$. Figure~\ref{fig1} shows the
stationary state occupation number of each node as a function of the degree. 
For $\delta\leq \delta_c$, the data follow the scaling form in 
Eq.~(\ref{m_scale}).  It is clearly seen that only the
hub is the condensed node at $\delta=0$.
At $\delta=0.8>\delta_c$, the condensation does not occurs and $m\sim
k^{1/\delta}$ for all nodes. 

In the context of transport on a SF network, one can avoid the
complete condensation if the transport capacity is designed to be greater than 
$p(n) = n^{\delta_c}$. The threshold value for $\delta$ is determined by
the network structure, i.e., $\delta_c = 1/(\gamma-1)$. 
When the complete condensation occurs, the high degree nodes 
are flooded by macroscopic numbers of particles. 
The condensation may slow down the dynamics leading to congestion.
We will study the dynamic aspect of the condensation.

Although the ZRP is solvable as far as the stationary state properties are
concerned, little has been known for 
its dynamical properties~\cite{Grosskinsky03,Godreche03}. 
We investigate relaxational dynamics of the ZRP on the BA network numerically.
Initially particles are distributed uniformly over the network. Observed 
is the time evolution toward the stationary state.

We propose that the quantity
$I(t) \equiv \sum_{i=1}^L \left({n_i(t)}/{N}\right)^2$
be useful in studying the condensation dynamics. It will be called the inverse
participation ratio~(IPR).
Roughly speaking, the IPR is given by the inverse of the number of nodes 
whose occupancy is non-negligible. The IPR in the stationary state, 
$I_{\text{stat.}} = \lim_{t\rightarrow\infty}I(t)$, can serve as the
order parameter for the condensation, for it is nonzero only 
in the condensed states. 

\begin{figure}[t]
\includegraphics*[width=\columnwidth]{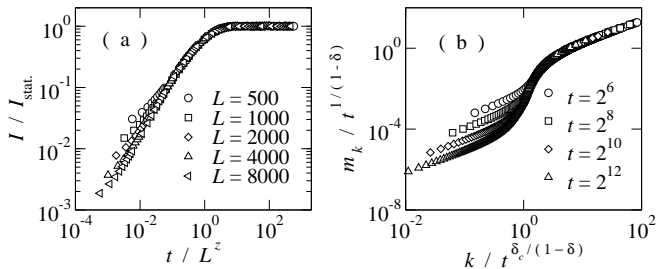}
\caption{Relaxation dynamics at $\delta=0.2$ in the BA network.
(a) Scaling plot of $I/I_{\text{stat.}}$ versus 
$t/L^z$ with $z=0.83$. (b) Scaling plot of
$m_k / t^{1/(1-\delta)}$ versus $k/t^{\delta_c / (1-\delta)}$ 
in the BA network of $L=10^6$ nodes.}
\label{fig2}
\end{figure}

We performed the Monte Carlo simulations on the BA networks.
In the condensed phase at $\delta\leq
\delta_c=1/2$, we found that the IPR converges to a finite
stationary state value $I_{\text{stat.}}$ and that 
the approach to $I_{\text{stat.}}$ becomes slower as $L$ increases.
It turns out that the relaxation time follows the power law
\begin{equation}\label{eq:tau}
\tau\sim L^z
\end{equation}
with a dynamic exponent $z$. Measuring the value of $\tau$ from
$I(t=\tau)=aI_{\text{stat.}}$ with a constant $a=0.8$,
we obtain that $z= 0.50$, 0.66, and 0.83 for 
$\delta=0.5$, 0.4, and 0.2, respectively.
Figure~\ref{fig2} (a) confirms the power-law scaling at $\delta=0.2$. 
On the other hands, the relaxation time does not grow algebraically in the 
non-condensed phase~\cite{Noh05}. 
Therefore we conclude that the dynamics becomes 
slow when the condensation occurs. 

We present a scaling argument for the power law $\tau\sim L^z$. 
In the periodic lattices the relaxation time is dominated by 
the process where macroscopic condensates at two sites merge 
into a single one~\cite{Grosskinsky03}. 
Similarly, we assume that it is dominated by the particle transfer process 
between two nodes 
with the largest occupancy, that is, the hub and the second largest hub.
Close to the stationary state, they are occupied by $\mathcal{O}(L)$
particles and a finite fraction of them should be transfered to the hub. 
At each time step, $\mathcal{O}(L^\delta)$ particles jump out of the nodes
and drift toward the hub.
We further assume that the drift time does not scale algebraically with $L$,
which might be justified by the small world property of the SF
networks~\cite{Albert02}. Under these assumptions, we obtain that
the transfer process takes $\mathcal{O}(L/L^\delta)$ time steps, 
hence $\tau \sim L^z$ with $z=1-\delta$. 
Note that it is consistent with the numerical results 
obtained from the IPR up to 10\%.  

As an evidence for the argument, we measured the mean occupation
number $m_k$ in the transient times $t\ll \tau$ in the BA network.
We find that at each time step $t$ there exists a degree scale 
$\tilde{k}=\tilde{k}(t)$ such that all nodes with 
$k\lesssim \tilde{k}$ behave as in the stationary state
in the smaller network with the maximum degree $\tilde{k}$.
Namely, $m_{k\leq \tilde{k}(t)}$ at time $t$ is given by the 
formula in Table~\ref{table1} with $k_{max}$ and $L$ 
being replaced with $\tilde{k}$ and $L'\sim \tilde{k}^{1/\delta_c}$, 
respectively. Then, the scaling argument can be validated by showing
that $t\sim L'^{(1-\delta)} \sim \tilde{k}^{(1-\delta)/\delta_c}$, i.e., 
$\tilde{k}(t) \sim t^{\delta_c / (1-\delta)}$. 
We confirm the scaling numerically with the scaling plot
of $m_k / L'$ versus $k/\tilde{k}$ at various times in Fig.~\ref{fig2}~(b), 
where all data collapse at $k\simeq \tilde{k}$.
We have also investigated the relaxation dynamics through the IPR and $m_k$
in various SF networks with different values of $\gamma$ and $\delta$,
and found that all numerical results are consistent with
$z=1-\delta$~\cite{Noh05}.

In summary, we have studied the condensation phenomena in the ZRP with the
jumping rate $p(n)=n^{\delta}$ on SF networks with the degree 
exponent $\gamma$. We have shown analytically 
that the complete condensation occurs when 
$\delta \leq \delta_c = 1/(\gamma-1)$. 
We have also shown numerically that the condensed phase with $\delta\leq
\delta_c$ has the slow relaxation dynamics, which is characterized by
the power-law scaling of the the relaxation time.

Acknowledgment: This work was supported by the KOSEF Grant 
No. R14-2002-059-01002-0 in the ABRL program.



\begin{references}
\bibitem{comment1} A generalization incorporating multiple particle jumping
         is studied recently in Ref.~\cite{Evans04}.
\end{references}
\end{document}